\documentstyle[12pt,fleqn,aasms4]{article}
\newcommand{\be}{\begin{equation}}
\newcommand{\ee}{\end{equation}}

\def\dol{{d_{\rm ol}}}
\def\dls{{d_{\rm ls}}}
\def\dos{{d_{\rm os}}}
\def\E{{{\rm E}}}

\def\kpc{{\rm kpc}}

\def\pc{{\rm pc}}

\def\kms{{\rm km}\,{\rm s}^{-1}}

\begin{document}

\title{A New Argument Against An Intervening Stellar Population Toward the
Large Magellanic Cloud}

\author{Andrew Gould\altaffilmark{1}}
\affil{Ohio State University, Department of Astronomy, 
174 West 18th Ave., Columbus, OH 43210}
\affil{E-mail: gould@astronomy.ohio-state.edu}

\begin{abstract}

Zaritsky \& Lin have claimed detection of an intervening population of stars 
toward the Large Magellanic Cloud (LMC) which, they believe, could account
for a substantial fraction of the observed microlensing events.  I show
that the observed time scales of these events imply that if such an
intervening population were composed of ordinary stars that gave rise to a 
significant fraction
of the microlensing events, then it could not be associated with 
the LMC.  I present two independent statistical arguments which together
essentially rule out such a chance alignment of unassociated structures.
On the other hand, if the intervening structure is associated with the LMC,
I show that of order half the mass in this structure is in substellar objects,
which would make it unlike any known stellar population.

\keywords{dark matter -- Galaxy: halo -- gravitational lensing 
-- Magellanic Clouds}
\end{abstract}

\section{Introduction}

	Zaritsky \& Lin (1997) found a concentration of stars approximately
0.9 magnitudes above the red clump (RC) in a color magnitude diagram (CMD)
of the Large Magellanic Cloud (LMC).  They suggested that this concentration
of stars might trace a
foreground population of ordinary stars, and this foreground population 
might be responsible for
a large fraction of the microlensing events seen toward the LMC by
Alcock et al.\ (1997a) and Aubourg et al.\ (1993).

	A number of workers raised a diverse set of objections to this
hypothesis.  Alcock et al.\ (1997b) showed that if the putative foreground
population lay within 33 kpc (i.e., 0.9 mag for an assumed LMC distance of
50 kpc), then it contained no detectable population
of RR Lyrae stars.  Beaulieu \& Sackett (1998) showed that a vertical red clump
(VRC), that is a vertical extension to the usual red clump, is a
typical feature of CMDs for populations of mixed age, and hence the presence
of a VRC did not necessarily indicate a foreground population.  Gallart
(1998) showed that such features are present in the Fornax and Sextans A
dwarf galaxies.  I argued (Gould 1998) that if such a foreground structure were
composed of tidal debris, then either it should have shown up in 
de Vaucouleurs' (1957) map of the LMC or it must have an anomalously high
mass-to-light ($M/L$) ratio to account for the microlensing events.  
Johnston (1998) showed that tidal debris from disrupted satellites would
give rise to unacceptably high star counts away from the LMC if it were to
account for the microlensing events seen toward the LMC.  Bennett (1998)
related the surface density of RC stars to the total surface mass density of 
their parent stellar distribution and showed that for typical stellar populations the density of the
VRC reported by Zaritsky \& Lin (1997) was too low by an order of magnitude
to account for the microlensing.

	Zaritsky et al.\ (1999) have addressed each of the objections in turn.
They said that it was possible to construct an initial mass function with 
a much higher ratio of total mass to RC stars than for the ``typical'' 
parameters advocated by Bennett (1998).  They argued that the foreground
population could be at 40 kpc, rather than the 33 kpc originally proposed by
Zaritsky \& Lin (1997), thus evading the constraint of Alcock et al.\ (1997b).
They pointed out that while certain star formation histories could well
explain the VRC as a feature of the LMC CMD as advocated by Beaulieu \&
Sackett (1998), such histories were not demanded by the available data, and 
indeed an independently constructed history yields only a small fraction
of the observed VRC.  They argued that Johnston's (1998) analysis does not
apply to tidal material from an SMC-LMC interaction or from a denser than
expected LMC halo.  Finally they quoted from de Vaucouleurs (1957) to make
it appear that he himself did not believe the outer isophotes of his map,
thereby apparently dispensing with my argument (Gould 1998).

	It is not my purpose here to critically examine all of these 
counter-arguments which would require a major investigation in its own
right.  Rather, I present a new argument against the hypothesis that
the VRC traces a significant lensing population.

\section{Transverse Speed of the Lenses}

	The speed of the lenses relative to the observer-source line of
sight, $v_\perp$, is related to the observed timescale of the events $t_\E$ by,
\begin{equation}
v_\perp = 190\,\kms \biggl({t_\E\over 40\,{\rm days}}\biggr)^{-1}
\biggl({M\over 0.25\,M_\odot}\biggr)^{1/2}
\biggl({\hat D\over 10\,\kpc}\biggr)^{1/2}, 
\qquad \hat D \equiv {\dol\dls\over\dos},
\label{eqn:veval}
\end{equation}
where $M$ is the mass of the lens, and $\dol$, $\dls$, and $\dos$ are the
distances between the observer, lens, and source.  This equation summarizes
the major difficulty in explaining the lenses as halo objects: if the
lenses were in the halo ($\hat D\sim 10\,\kpc$) and were 
substellar objects ($M<0.08\,M_\odot$), then to produce events with the
observed time scales ($t_\E\sim 40\,$days), they must be moving with
typical speeds $v_\perp\la 110\,\kms$, which is more than a factor of two smaller
than the speeds expected from halo dynamics.  Hence, if they are in the
halo, they are not made of hydrogen: substellar objects would be moving too
slowly while stellar objects made of hydrogen would burn and be visible
(e.g. Gould, Flynn, \& Bahcall 1998 and references therein).

	Equation (\ref{eqn:veval}) can also be used to draw significant
conclusions about the putative foreground structure claimed by Zartisky \&
Lin (1997).  If this structure is composed of ordinary stars
($M\sim 0.25\,M_\odot$), and if it lies 0.9 mag  (17 kpc) in front of the
LMC ($\hat D=11\,\kpc$), then it must be travelling at $v_\perp\sim 200\,\kms$
relative to the line of sight to LMC.  This is approximately what would
be expected for a random object travelling through the Galactic halo
such as a dwarf galaxy or tidal debris from a disrupted dwarf (Zhao 1998).  
However, it is substantially too high for material associated with the LMC.

	Hence, if the claimed foreground structure is truly responsible
for a substantial fraction of the microlensing events, then there are two 
possibilities: either the foreground structure is not associated with the LMC,
or it is associated but is composed of objects that are substantially lighter
than the mass of typical stars, $M\sim 0.25\,M_\odot$.  I examine these two
possibilities in turn, beginning with the hypothesis of a chance alignment
of a structure unassociated with the LMC.

	The a priori probability of such an alignment is incredibly small.
Recall from Gould (1998) that the surface mass density required to explain
the observed microlensing optical depth, $\tau\sim 2.9\times 10^{-7}$, is
\begin{equation}
\Sigma = 47\,{\tau\over 2.9\times 10^{-7}}\biggl({\hat D\over 10\,\kpc}
\biggr)^{-1} M_\odot\,\pc^{-2}.
\label{eqn:sigmaeval}
\end{equation}
For a Milky-Way disk $M/L=1.8$, this corresponds to about 
$V=22.8$ mag arcsec$^{-2}$.  By comparison, the central surface brightnesses of
the Sculptor and Sextans dwarf spheroidal galaxies are respectively
23.7 and 26.1 mag arcsec$^{-2}$ (Mateo et al.\ 1991).  
Moreover, the core radii of these galaxies
are only 9 and 15 arcmin respectively (Mateo et al.\ 1991), much smaller
than the several square degrees required to account for the microlensing
events.  The fraction of the high-latitude sky covered by dwarf galaxies
of even these low surface brightnesses is well under $10^{-4}$.
(Note that the a priori probability of a structure {\it associated with the 
LMC} would not be affected by this argument.) \ \ 

\section{Expected Radial-Velocity Difference}

	However, it is not entirely appropriate to apply a priori
statistical arguments to the presence of a dwarf galaxy in front of
the LMC.  The fact is that microlensing events have been discovered toward
the LMC and {\it all} the explanations offered so far are a priori unlikely.
If evidence is produced for an intervening stellar population {\it after the
detection of the microlensing events} (e.g. Zaritsky \& Lin 1997), 
then the low a priori probability for such a population carries less weight.

	Nevertheless, this putative detection brings with it the means for
an additional, truly a priori test.  If the intervening population is
not associated with the LMC, then
its radial motion relative to the LMC should be a random value drawn from
a distribution characteristic of Galactic satellites.  On the other hand, 
if the VRC is actually composed of LMC stars, the two radial velocities 
should be consistent.   

I find that the
root-mean-square galactocentric radial velocity of 16 satellite galaxies and
distant globular clusters at high latitude
(422-213, 
 AM1,
 Carina,
 Draco, 
 Fornax,  
 LMC,
 LeoI,   
 LeoII,   
 NGC2419,
 Pal3, 
 Pal4,  
 Pal14,
 Sculptor,
 Sextans,
 SMC, and
 UMi) 
is $\sigma_{\rm sat}= 86\,\kms$.
Zaritsky et al.\ (1999) have measured the mean radial velocities
of the two populations and find a difference,
\begin{equation}
\Delta {\bar v}={\bar v}_{VRC} - {\bar v}_{RC} = 5.8 \pm 4.9\,\kms,
\label{eqn:vrcomp}
\end{equation}
where (in contrast to Zaritsky et al.\ 1999) I am quoting $1\,\sigma$ errors.
This confirms the results of Ibata, Lewis, \& Beaulieu (1998) based on a
smaller sample
and is just what one would expect if the VRC stars were part of the LMC.
However, the a priori probability that the two populations would be this close
if they were not associated is 
\begin{equation}
p \sim {2\Delta\bar v\over (2\pi)^{1/2}\sigma_{\rm sat}}\exp
\biggl(-{v_{\rm LMC}^2\over 2 \sigma_{\rm sat}^2}\biggr)\sim 4\%,
\label{eqn:prob}
\end{equation}
where $v_{\rm LMC}=73\,\kms$ is the galactocentric radial velocity of the LMC.

	In brief, there are two distinct statistical arguments against the VRC
being a foreground structure that is not associated with the LMC: first it is
unlikely ($<10^{-4}$) that such a structure would happen to be
aligned with the LMC, and second it is unlikely (4\%) that it
would have a radial velocity consistent with that of the LMC.  Together,
these two arguments effectively rule out this possibility.

\section{Mass Scale of Foreground Population}

	I therefore turn to the second possibility discussed in \S\ 2, that
the masses of the foreground objects are substantially smaller than those
of typical stars.  To investigate this possibility, it is first necessary
to estimate the typical transverse velocities of populations that {\it are}
 associated with the LMC.  For a foreground population with the same
space velocity as the LMC but lying a distance $\dls$ in front of it, the
transverse speed of the lens population relative to the Earth-LMC line
of sight is
\begin{equation}
v_{\perp,\rm bulk} = \mu_{\rm LMC}\dls = 109\,\kms\,{\dls\over 17\,\kpc}
\label{eqn:mulmc}
\end{equation}
where $\mu_{\rm LMC} = 1.35\,\rm mas\,yr^{-1}$ is the proper motion of the
LMC (Jones, Klemola, \& Lin 1994).  To find $v_\perp$, this bulk motion
must be added to any peculiar motion of the sources and lenses relative
to the assumed common space motion.  The internal dispersion of the VRC
is small, $18\,\kms$ (Zaritsky et al.\ 1999), and therefore can be ignored.
The bulk motion of the foreground structure relative to the LMC must also
be small to evade the the arguments given in the last two sections about
unassociated structures, so this will also be ignored.  However, the LMC
sources are rotating at $\sim 70\,\kms$ and this must be included.  It
should first be multiplied by the projection factor $\dol/\dos$ and then
added in quadrature (because the microlensing observations cover a sufficiently
large part of the LMC that all directions of motion are effectively covered).
Hence, at $\dls=17\,\kpc$, the expected $v_\perp = 119\,\kms$.  Inserting
this value into equation (\ref{eqn:veval}) yields a typical mass
$M\sim 0.09\,M_\odot$ at $\dls=17\,\kpc$.  At $\dls=10\,\kpc$, the
same argument yields $M\sim 0.06\,M_\odot$.  Another possibility for 
a foreground population at $\dls=10\, \kpc$ is that it is a bound satellite 
orbiting about the LMC at $\sim 70\,\kms$.  However, this scenario leads
to essentially the same mass, $M\sim 0.07\,M_\odot$.  For distances
$\dls\ll 10\,\kpc$, it is no longer plausible that the foreground population
would give rise to the observed VRC which peaks 0.9 mag brighter than the RC.
Thus, $M\sim 0.08\,M_\odot$ is a robust estimate of the characteristic
mass of the putative foreground population.

	Since $M\sim 0.08\,M_\odot$ is approximately where hydrogen burning 
begins, this result implies that of order half the mass in the putative
foreground structure lies below the hydrogen burning limit.  While this
is possible in principle, it should be noted that in the solar neighborhood,
substellar objects account for only about 1/6 of the total stellar and
substellar mass (Holmberg \& Flynn 1999 and references therein).

{\bf Acknowledgements}:  
I thank B.S. Gaudi for a careful reading of the manuscript.
This work was supported in part by grant AST 97-27520 from the NSF.



\clearpage

\begin{references}
\reference{alc} Alcock et al.\ 1997a, \apj, 486, 697 
\reference{alc} Alcock et al.\ 1997b, \apj, 490, 59  
\reference{Aubourg} Aubourg., et al.\ 1993, Nature, 365, 623
\reference{BS} Beaulieu, J.-P., \& Sackett, P.\ D.\ 1998, \aj, 116, 209
\reference{alc} Bennett, D.\ 1998, \apj, 493, L79
\reference{deV} de Vaucouleurs, G.\ 1957, \aj, 62, 69
\reference{gall} Gallart, C.\ 1998, \apj, 43
\reference{gtwo} Gould, A.\ 1998, \apj, 499, 728
\reference{gfb} Gould, A., Flynn, C., \& Bahcall J.N.\ 1998, \apj, 503, 798
\reference{hf} Holmberg, J.\ \& Flynn, C.\ 1999, \mnras, submitted
\reference{jkl} Jones, B.F., Klemola, A.R., \& Lin D.N.C.\ 1994, \aj, 107, 
1333
\reference{jstn} Johnson, K. V.\ 1998, \apj, 495, 297
\reference{ilb} Ibata, R.A., Lewis, G.F., \& Beaulieu, J.-P. 1998, \apj, 509,
L29
\reference{mateo} Mateo, M., Nemec, J., Irwin, M., \& McMahon, R.\ 1991,
\aj, 101, 892
\reference{zarit} Zaritsky, D., \& Lin, D.\ N.\ C. 1997, \aj, 114, 2545
\reference{zar2} Zaritsky, D., Shectman, S.A., Thompson, I., Harris, J.,
\& Lin, D.\ N.\ C. 1999, \aj, 117, 000
\reference{zhao} Zhao, H.\ 1998, \mnras, 294, 139
\end{references}
\end{document}